\documentclass[10pt, final, comsoc,twocolumn]{IEEEtran}
\IEEEoverridecommandlockouts
 \pdfoutput=1
\usepackage[cmex10]{amsmath}
\usepackage{amsfonts}
\usepackage{amssymb}
\usepackage{amscd}
\usepackage{float}
\usepackage{cite}
\usepackage{mathrsfs}
\usepackage{afterpage}
\usepackage{graphicx}
\usepackage{epstopdf}
\usepackage{caption}
\usepackage{subcaption}
\usepackage{color}

\newcommand{\beq}{\begin{equation}}
\newcommand{\eeq}{\end{equation}}
\newcommand{\beqa}{\begin{eqnarray}}
\newcommand{\eeqa}{\end{eqnarray}}
\newcommand{\nn}{\nonumber\\}

\title{Half-Duplex or Full-Duplex Communications: Degrees of Freedom Analysis under Self-Interference}

\author{Nirmal~V.~Shende,~\IEEEmembership{Student Member,~IEEE,}  
Ozgur~Gurbuz,~\IEEEmembership{Member,~IEEE,}
 and Elza~Erkip~\IEEEmembership{Fellow,~IEEE,}
\thanks{N.~V.~Shende is with the School of Electrical and Computer Enginnering, Cornell University, Ithaca, NY 14853, USA (email:nvs25@cornell.edu).}
\thanks{ O.~Gurbuz is with the Faculty of Engineering and Natural Sciences, Sabanci University, Istanbul 34956, TURKEY (email:ogurbuz@sabanciuniv.edu).}
\thanks{E. Erkip is with  the Department of Electrical and Computer Engineering, New York University, Tandon School of Engineering, Brooklyn, NY 11201, USA (email:elza@nyu.edu).}
\thanks{This work was supported in part by NSF Grant No 1527750 and TUBITAK Grant No 113E222. 
The material in this paper was presented in part, with earlier results, at the Conference on Information Sciences and Systems (CISS 2013)~\cite{CISS}.}
}
\begin{document}
\maketitle

\begin{abstract}
In-band full-duplex (FD) communication provides a promising alternative to half-duplex (HD) for wireless systems, due to increased spectral efficiency and capacity. In this paper, HD and  FD radio implementations of two way, two hop and two way two hop communication are compared in terms of degrees of freedom (DoF)  under a realistic residual self-interference (SI) model. DoF analysis is carried out for each communication scenario for HD, antenna conserved (AC) and RF chain conserved (RC) FD radio implementations. The DoF analysis indicates that for the two way channel,  the achievable AC FD with imperfect SI cancellation performs strictly below HD, and RC FD DoF trade-off is superior when the SI can be sufficiently cancelled. For the two hop channel, FD is better when the relay has large number of antennas and enough SI cancellation. For the two way two hop channel, when both nodes require similar throughput,  the achievable DoF pairs for FD do not outperform HD. FD still can achieve better DoF pairs than HD, provided the relay has sufficient number of antennas and SI suppression. 
\end{abstract}
\section{Introduction}
In almost all  networks, a communicating device has a dual task of
reception and transmission of data. This is commonly achieved via
half-duplex (HD) operation, where the channel is time shared between
transmission and reception, so that a node can either transmit or
receive at a given time. Full-duplex (FD) operation provides a
promising alternative, where both of these activities are implemented
simultaneously. However, an FD node suffers from high amount of
self-interference (SI), since typically the transmitted signal is
about 100 dB stronger than the received signal. Recently
FD has gained considerable interest due to
promising results on practical implementations
\cite{Duarte10,Knox12,choi,Khojastepour,Jain11}, as can be seen in
the recent review article \cite{Sabharwal_Survey} and references
therein.

Ideally, FD implementation uses the channel for transmitting and
receiving simultaneously, and hence it is likely to give higher
throughput. On the other hand, FD requires hardware resources, such
as antennas to be divided between transmission and reception, in
order to accomplish this with as little SI as possible. However, since SI cannot be suppressed completely, the residual SI reduces the received signal-to-interference-noise ratio (SINR), resulting in reduced data rates. Hence, how much improvement can FD communication in the presence of SI can provide over HD, considering similar hardware resources is an important question that needs to be investigated thoroughly for viability of FD.
In order to address this problem, in this paper we compare wireless HD and FD communication in three communication scenarios, two way, two hop (relaying), and two way two hop (two way relaying) systems, illustrated in Figures \ref{fig:TWC}-\ref{fig:TWRC2}, from degrees of freedom (DoF)  point of view. The system models considered in this paper arise naturally in modern communication scenarios, such as cellular, WiFi, mesh or ad-hoc networks, which would particularly benefit from FD implementations.

One of the challenges in analytical study of the FD systems is the
modeling of the residual SI. The model should be accurate, so that
it captures the effect of SI, and also simple enough, so that it is
useful for analysis and design. Some works assume constant increase
in the noise floor due to SI \cite{Bharadia,Jain11}. However, it is
reasonable to expect that SI will depend on the transmit power. Other
works assume linear increase in SI with transmit power
\cite{Day12,Riihonen11,Ng}, but this model fails to capture the
effect, in which increased transmission power actually enhances SI
suppression, since a better estimate of the SI signal is obtained.
In our analysis in this paper, we use the experimentally validated
SI model from \cite{SImodel}, which shows that average residual SI
power after cancellation can be modeled as proportional to $P^{1-\lambda}$, where
$0\le\lambda\le 1$ is a constant that depends on the transceiver's
ability to mitigate SI and $P$ is the transmit power. This model, also used in \cite{Rodriguez2}, not only captures the effect of the practical SI cancellation mechanisms employed, but it is analytically tractable as well. Furthermore, it generalizes all other models used in the literature.

In order to provide a fair comparison of FD and HD implementations,
it is important to keep the hardware resources fixed. For this
purpose, we follow two approaches as in \cite{Sanaz}: For each node,
we either keep the total number of antennas or we keep the total
number of RF chains of FD mode the same as that of HD mode,
considering \emph{antenna conserved} (AC) and \emph{RF chain
conserved } (RC) implementations of FD, respectively. The AC FD scenario is motivated by the recent FD implementations \cite{Duarte10,Jain11}; the notion of keeping the number of RF chains equal is also reasonable from a practical perspective, since RF chains are the components that dominantly increase the total cost of a radio \cite{Aggarwal12}.

In this paper, considering the three                                                    scenarios, namely two way, two hop and two way two hop communication, under realistic SI and hardware constraints,
we pursue the high-SNR DoF
analysis \cite{Tse} for comparison of the performances of HD and FD modes. The
DoF metric admits simple analytical characterization facilitating
the comparisons. Our analysis in this paper not only provides the guidelines for selection of HD or FD mode for the considered scenarios, but it also sets forth the basic models  for future studies for more complex scenarios.  Our main observations can be summarized as follows:
\begin{itemize}
\item For the two way channel (Figure \ref{fig:TWC}), we show that in presence of SI ($\lambda<1$), the FD DoF region, which shows the simultaneously achievable DoF pairs by both users for the AC scenario lies strictly inside the HD trade-off. For the RC scenario, however, with ``good'' SI suppression (typically $\lambda>0.75$), FD can achieve certain DoF pairs which are not achievable by the HD implementation. 

\item For the two hop channel (a relay channel without a direct link between the source and destination) as shown in Figure \ref{fig:OWRC}, we compare the FD and HD DoFs for the symmetric case (when both source and destination have equal number of antennas) and the asymmetric case (when source has a single antenna, and destination has multiple antennas). We find that, for given number of source and destination antennas, and SI parameter $\lambda$, the FD implementation outperforms HD if the relay has sufficient number of antennas, otherwise HD is better. Number of antennas required at the relay for this crossover is lower for the RC scenario, than that of the AC scenario, and depends on the SI mitigation level $\lambda$ and the number of source and destination antennas. When the number of antennas at each node is fixed, then there exists threshold value of $\lambda$, below which HD achieves higher throughput than FD. 

\item For the two way two hop channel (two way relay channel without a direct link between the communicating nodes, as shown in Figures \ref{fig:TWRC1} and \ref{fig:TWRC2}), with only the relay having FD capability, in both AC and RC scenarios, if the symmetric DoF is to be maximized, then generally HD performs better than FD. For the asymmetric case however, provided the SI suppression is high enough (in terms of $\lambda$), FD can achieve certain DoF pairs which are not achievable by the HD. These pairs generally correspond to the extreme asymmetric DoF, when one node's DoF requirement is significantly higher than the other one. 
\end{itemize}

\subsection{Related Literature}

Recently, there has been a significant body of work on FD communications, and here, we briefly summarize the most relevant papers. In \cite{Ju11}, the achievable sum rates in a two way channel for FD and HD are compared assuming perfect SI cancellation for AC implementation. Reference \cite{estimation} compares the FD and HD
two way channel in the presence of channel estimation errors, and
depending on the level of SI and the channel estimation errors, an
outer bound for the region over which FD is better than HD is
provided. An outage analysis for FD two way communication under
fading can be found in \cite{Arifin}. In \cite{Comtel},  results on the sum rate performance of two way HD and FD communication are presented considering the FD implementations from \cite{Sanaz}, which are optimistic for the RC FD implementation with larger number of antennas. In \cite{FDMIMO}, a study on FD Multiple Input Multiple
Output (MIMO) system is presented, basically showing how a common
carrier based FD radio with a single antenna, as in \cite{Knox12},
can be transformed into a common carrier FD MIMO radio.

In \cite{Pashazadeh}, two hop communication is studied with channel
estimation errors in the presence of loop-back interference in order
to come up with capacity cut-set bounds for both HD and FD relaying.
An effective transmission power policy is proposed for the relay to
maximize this bound, and performance of FD relaying with optimal
power control is compared with HD relaying. Two hop communication in
a cellular environment is investigated in \cite{goyal_improving},
where a hybrid scheduler that is capable of switching between HD and FD in an opportunistic fashion is proposed, for maximizing the system throughput. Reference \cite{Riihonen11} has shown that, in order to control the SI, the relay should employ power control and the proposed relaying scheme allows to switch between HD and FD modes in an opportunistic fashion, while transmit power is adjusted to maximize spectral efficiency. In \cite{CISS},  results in the relaying scenario are presented, comparing FD and HD relaying under
the empirical residual SI model from \cite{SImodel}. In that work, power control is used asymptotically, so that the relay scales its power with respect to the source to achieve maximum DoF, when the relay operates in decode and forward mode. Similar asymptotic power control was also observed to give higher DoF in amplify and forward in mode in
\cite{Rodriguez2}. In \cite{Alves}, two way relaying HD and FD
systems are analyzed, where source and destination nodes are assumed
to hear each other. A survey on FD relaying can be found in
\cite{FDR_Survey}.

The current literature does not contain a detailed investigation of the DoF  analysis for the three communication scenarios under realistic SI and hardware constraints, as considered in this paper. In most of the existing literature, a specific self-interference (SI) model is used, and the HD and FD performance is compared
for that SI model. Furthermore, the SI model used either assumes SI power scales linearly
with transmit power, or SI is taken simply as an increase in the noise floor. In this paper, we
use a generalized and experimentally validated SI model that incorporates and generalizes
both scenarios and compare the HD and FD performance. We study the DoF of three
important building blocks of a wireless network: two way communication, single hop and
two way two hop. These channel models and the analysis illustrate the fundamental benefits and limitations of using FD in typical wireless scenarios. 

\subsection{Paper Organization}

The rest of the paper is organized as follows: Section \ref{sec:system_model} describes the considered three system models for two way, two hop and two way two hop communication, together with channel, FD implementation and SI cancellation models. In Sections \ref{sec:TWC}-\ref{sec:TWTHC}, the DoF analysis is presented with detailed comparisons and discussions of the HD and FD implementations of the three system models. Section \ref{sec:conclusions} involves our concluding remarks.

\section{System Models}
\label{sec:system_model} In the following, we describe the three different scenarios, two way, two hop and two way two hop communication, in which FD can be implemented. We start by providing a wireless channel model between two nodes, as a generalized point-to-point channel model that will be used throughout the paper. Then, for each communication model, we present the information flow for both HD and FD implementations.

\subsection{Generic Channel Model Between Two Nodes}
Consider a scenario where node $A$ is transmitting to node $B$, where node $B$ is operating either in HD mode or FD mode depending upon scenario being investigated. Let $P_{A}$ denote the average transmit power at node $A$, $\sigma_{B}^2$ is the average power of the AWGN at node $B$. Nodes are assumed to have multiple antennas with $\mathbf{H}_{AB}$ denoting the channel matrix between nodes $A$ and $B$. We assume Rayleigh fading channel, so the entries of $\mathbf{H}_{AB}$ are taken as independent and identically distributed circularly symmetric complex Gaussian random variables with unit variance \cite{Goldsmith}. Channel state information is assumed to be available only at the receiver. The size of the channel matrices depends on the number of the transmit and receive antennas employed at the nodes. Then, the received signals at node $B$ is
\begin{align*}
\mathbf{y}_B &= \frac{1}{\sqrt{K}}\mathbf{H}_{AB}\mathbf{x}_{A}+\mathbf{w}_B+\mathbf{i}_{B}.
\end{align*}
Here, $\mathbf{x}_{A}$ denotes the vector of the transmitted symbol,
$\mathbf{w}_B$ denotes the noise term, and $\mathbf{i}_{B}$ is the
SI term if node $B$ is operating in FD mode. We  assume entries
of $\mathbf{i}_{B}$ are Gaussian distributed with variance equal to
average SI power. This assumption makes analysis tractable and can
be viewed as the worst case scenario, since Gaussian distribution
gives worst case capacity \cite{Gaussian_Noise1,Gaussian_Noise2}.
Clearly, in the case of HD, this term is set to zero. Expected value of $i_B$ will be denoted as $I_B$. Finally, $K$ is the
parameter that characterizes the path loss between nodes. The SINR
at the receiver is, \beq
\Gamma_{AB}=\frac{P_{A}}{K\left(\sigma_{B}^2+ I_{B}(P_B)\right)},
\label{eq:SNR} \eeq where we have explicitly showed the dependence
of the average SI, $I_{B}$ on $P_{B}$, the transmit power of $B$ according to
\cite{SImodel}. Details of the SI model will be described in Section~\ref{subsec:SI_model}. Then, assuming that node $A$ transmits using $N_A$ antennas and
node $B$ receives using $N_B$ antennas, the average achievable rate
$R_{AB}$ is \cite{Tse}
\begin{align*}
R_{AB}=\mathbb{E}\left[\log\det\left(\mathbf{I}+\frac{\Gamma_{AB}}{N_A}\mathbf{H}_{AB}\mathbf{H}_{AB}^*\right)\right].
\end{align*}
Degrees of Freedom (DoF) analysis characterizes the achievable rate at high SNR. For a point to point MIMO AWGN Rayleigh fading channel with $N_A$ antennas at $A$  and $N_B$ antennas at $B$, the largest DoF is given by\cite{Tse}
\begin{align*}
\text{DoF}_{AB}=\lim_{P_A\to\infty}\frac{R_{AB}}{\log(P_A)}=\min\left(N_B,N_A\right).
\end{align*}

\subsection {Communication Scenarios}
\label{system_models}
\subsubsection{Two Way Channel}

\begin{figure}[]
\centering
\begin{subfigure}[]{\linewidth}
  \centering
  \includegraphics[scale=0.30]{TWC_HD_AB.png}
  \caption{HD communication with $A$ transmitting to  $B$}
  \label{fig:TWC_HD_AB}
\end{subfigure}\\
\begin{subfigure}[]{\linewidth}
  \centering
  \includegraphics[scale=0.30]{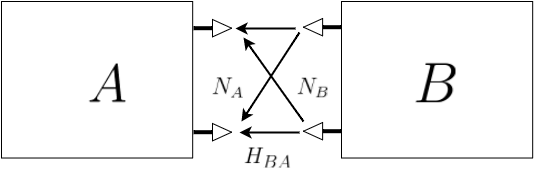}
  \caption{HD communication with $B$ transmitting to  $A$}
  \label{fig:TWC_HD_BA}
\end{subfigure}
\begin{subfigure}[]{\linewidth}
  \centering
  \includegraphics[scale=0.30]{TWC_FD.png}
  \caption{FD communication, dotted arrows indicate SI.}
  \label{fig:TWC_FD}
\end{subfigure}
\caption{Two way channel}
\label{fig:TWC}
\end{figure}

Two way communication channel was introduced by Shannon in
\cite{shannon1961}. Here, we consider a two way wireless channel,
where node $A$ and $B$ have $N_A$ and $N_B$ antennas respectively,
and wish to communicate with one another.  This channel, for example
may model a WiFi router communicating with a wireless device, which
is simultaneously uploading and downloading data. In the HD mode,
the nodes time share the wireless medium, taking turns transmitting
as shown in Figures \ref{fig:TWC_HD_AB} and \ref{fig:TWC_HD_BA}. In
this case, nodes use all of their antennas either for transmission
or for reception. If both $A$ and $B$ are FD capable, then they can
use the channel simultaneously for transmission and reception, as
shown in Figure \ref{fig:TWC_FD}. Here, $t_A$ and $r_A$ denote the
number of transmit and receive antennas at node $A$, respectively.
Similarly, $t_B$ and $r_B$ denote the number of antennas at node
$B$. Dotted arrows in each direction represent the SI channels. The
choice of $t_A, r_A, t_B$ and $r_B$ based on hardware constraints
will be discussed in Section \ref{sec:hardware_resource}.

\subsubsection{Two Hop Channel}
\label{sec:OWTHC_model}

In this scenario, node $A$ communicates with node $B$ through a
relay node, $R$. We assume that there is no direct link between
nodes $A$ and $B$, hence the relay assists in forwarding the packets
from $A$ to $B$. The relay is assumed to operate according to decode
and forward protocol, \cite{DaF}. Nodes $A$ and $B$ have $N_A$ and $N_B$
antennas, respectively. Total number antennas employed in $R$ in the
HD mode is denoted by $N_R$. When the relay operates in FD mode, then $t$
and $r$ denote the number of transmit and receive antennas
respectively.

\begin{figure}[]
\centering
\begin{subfigure}{\linewidth}
  \centering
  \includegraphics[scale=0.33]{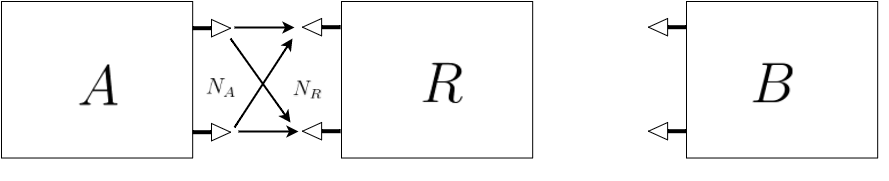}
  \caption{First phase of HD relaying with  $A$ transmitting to  $R$}
  \label{fig:OWRC_HD_AR}
\end{subfigure}
\begin{subfigure}{\linewidth}
  \centering
  \includegraphics[scale=0.25]{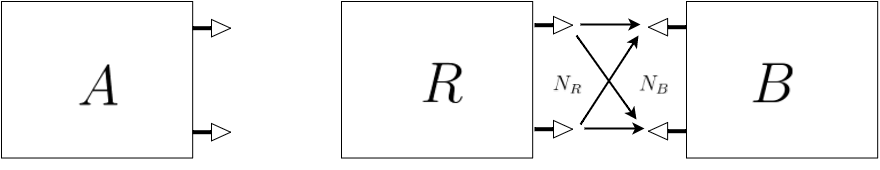}
  \caption{Second phase of HD relaying with  $R$ transmitting to  $B$}
  \label{fig:OWRC_HD_RB}
\end{subfigure}
\begin{subfigure}{\linewidth}
  \centering
  \includegraphics[scale=0.25]{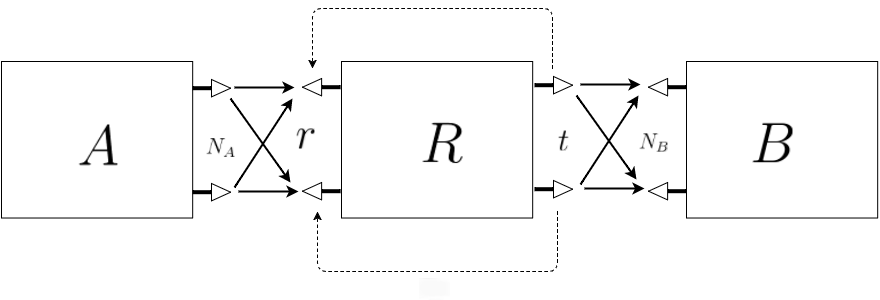}
  \caption{FD relaying with  $A$ transmitting to  $R$, and  $R$ transmitting to  $B$, \\dotted arrows indicate SI}
  \label{fig:OWRC_FD}
\end{subfigure}
\caption{Two hop channel}
\label{fig:OWRC}
\end{figure}

When the relay is in HD mode, the information flow takes place in two
phases: First, $A$ transmits to $R$ as shown in Figure
\ref{fig:OWRC_HD_AR}, and then $R$ decodes the packets and forwards
to $B$, as shown in Figure \ref{fig:OWRC_HD_RB}. In the case of FD
relaying, $R$ can receive from $A$ and simultaneously transmit to
$B$, as shown in Figure\ref{fig:OWRC_FD}. It allocates its resources
(antennas or RF chains) so as to increase the data rate from $A$ to
$B$.
\subsubsection{Two Way Two Hop Channel}

This channel models a two way relay channel without a direct link
between communicating nodes. Here, two nodes $A$ (with $N_A$
antennas) and $B$ (with $N_B$ antennas) wish to communicate with
each other through a relay  $R$ (with $N_R$ antennas in HD mode).
Only $R$ is assumed to have FD capability and uses $t$ antenna for transmission and $r$ antenna for reception in FD mode. A motivating example for
such scenario is two stations on earth communicating via a
satellite, with no direct link between the stations.

\begin{figure}[h]
\centering
\begin{subfigure}{\linewidth}
  \centering
  \includegraphics[scale=0.40]{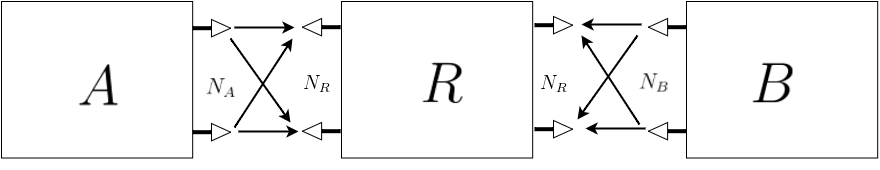}
  \caption{MAC phase, nodes $A$ and $B$ transmitting to $R$}
  \label{fig:TWRC_HD_MAC}
\end{subfigure}
\begin{subfigure}{\linewidth}
  \centering
  \includegraphics[scale=0.40]{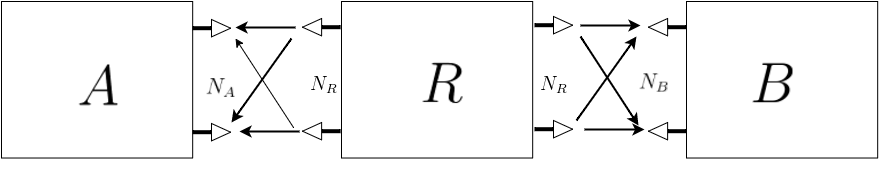}
  \caption{BC phase, $R$ transmitting to $A$ and $B$}
  \label{fig:TWRC_HD_BC}
\end{subfigure}
\caption{HD two way two hop channel} \label{fig:TWRC1}
\end{figure}

When $R$ is operated in HD mode, we consider an effective communication
strategy, such as \cite{4418498,katti_embracing,Zhang06}, which
takes place in two phases, as shown in Figures \ref{fig:TWRC_HD_MAC}
and \ref{fig:TWRC_HD_BC}. During the first phase, also known as the
multiple access (MAC) phase, nodes $A$ and $B$ simultaneously
transmit to $R$. During second phase, called broadcast (BC) phase,
$R$ simultaneously transmits to $A$ and $B$, and both nodes can
extract their desired signal by the virtue of analog coding
techniques.

\begin{figure}[h]
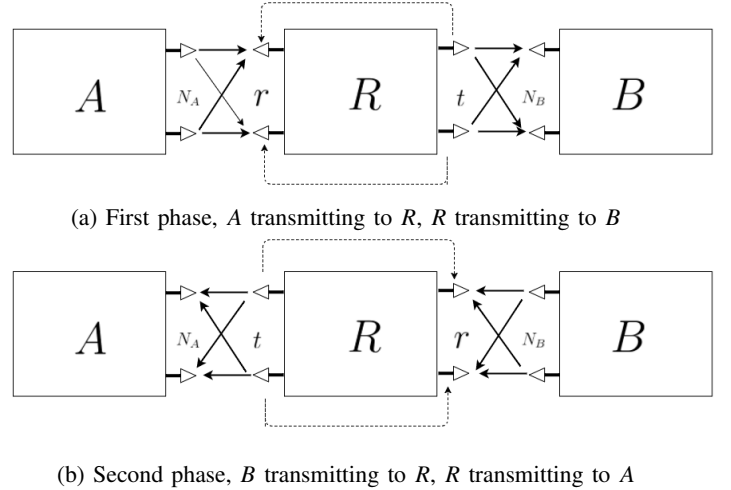

\centering
\begin{subfigure}{\linewidth}
  \centering
  \includegraphics[scale=0.30]{TWRC_FD_AB.png}
  \caption{First phase, $A$ transmitting to $R$, $R$ transmitting to $B$ }
  \label{fig:TWRC_FD_First}
\end{subfigure}
\begin{subfigure}{\linewidth}
  \centering
  \includegraphics[scale=0.30]{TWRC_FD_BA.png}
  \caption{Second phase, $B$ transmitting to $R$, $R$ transmitting to $A$}
  \label{fig:TWRC_FD_Second}
\end{subfigure}
\caption{FD two way two hop channel, dotted arrows indicate SI} \label{fig:TWRC2}
\end{figure}

For the FD case, only $R$ is assumed to have FD capability, and two
way FD communication occurs in two phases, as shown in Figures
\ref{fig:TWRC_FD_First} and \ref{fig:TWRC_FD_Second}. During the
first phase node $A$ transmits to $B$ via $R$, and since $R$ is FD,
it can receive from node $A$ and transmit simultaneously to $B$.
During the second phase, direction of information flow is reversed,
as node $B$ transmits to $A$ via $R$.


\subsection{SI Cancellation Model}
\label{subsec:SI_model}
The major challenge of FD communication is SI cancellation. As
discussed in detail in \cite{3models}, the simplest SI cancellation
technique is the passive one, obtained by the path-loss due to the
separation between the transmit and receive antennas. More
sophisticated active techniques, namely, analog cancellation and
digital cancellation reduce the self-interference further. In analog
cancellation, the FD node uses additional RF chains to estimate the
channel between the transmitting and receiving antennas and then to
subtract the interfering signal at the RF stage. In the digital
cancellation, the self-interference is estimated and canceled in the
baseband. Despite consecutive application of these three
cancellation techniques, SI cannot be completely eliminated. In
\cite{SImodel}, the average power of the residual SI is experimentally modeled as
\beq
 {I} =\frac{P_T^{\left(1-\lambda \right)}}{\beta \mu^{\lambda}}.
\label{eq:SImodel1} \eeq Here, $P_T$ denotes the transmission power
of the FD node. $\beta$, $\mu$, and $\lambda$ are the system
parameters which depend on the cancellation technique employed, with
$0 \leq \lambda \leq 1$. Note that, $\lambda=1$ corresponds to
increased noise floor SI model used in the literature and $\lambda =0$ corresponds to SI power scaling linearly with the transmit power.

This SI model was obtained for a FD transmitter with a single receive and single transmit
antenna. In the case of a multiple transmit and receive antenna FD terminal, we  could implement transmit precoding and receive processing to further mitigate the SI. At each receive antenna, this would at most increase $P_T$ in (\ref{eq:SImodel1}) by a factor of $t$ (number of transmit antennas), which for the purpose of a DoF analysis, would have the same effect as (\ref{eq:SImodel1}). Hence in this paper we continue to use (\ref{eq:SImodel1}) to model the average  residual SI  power per receive antenna.

A natural question then arises: can DoF can be improved by using such transmit and linear precoding and receive processing? In other words, is a DoF analysis based on the model in (\ref{eq:SImodel1}) unnecessarily pessimistic?  We believe that is not the case since the SI channel matrix is generally full rank~\cite{softnull}. As reported in~\cite{Shojaeifard}, unless the transmission is carried out in the null space of the SI channel, the SI power continues to scale with transmit power $P_T$,  leading to the model in (\ref{eq:SImodel1}) for the purposes of a DoF analysis. On the other hand,  if the encoder attempts to transmit in the null-space of the SI channel matrix, it would lead to loss in DoF since the SI matrix is full rank.  Moreover, any such strategy would require very accurate estimates of the SI channel. We illustrate this for the case of the two way channel in Section~\ref{subsec:comp_twc}, where we show that in obtaining the DoF, the model in (\ref{eq:SImodel1}) is sufficient. Another advantage of the model in (\ref{eq:SImodel1}) is that it simply defers all SI mitigation
to hardware and does not use  any SI channel knowledge or SI management  strategy while designing the transmit signal. We must add, however, that obtaining converse results, which show that no multi-antenna processing would improve the DoFs beyond the ones obtained in this paper,   is reasonably arduous, as the SI is in general non-Gaussian.
\subsection{Hardware Resources in HD and FD}
\label{sec:hardware_resource}
\begin{figure}[htp]
\centering
\begin{subfigure}{\linewidth}
  \centering
  \includegraphics[scale=0.28]{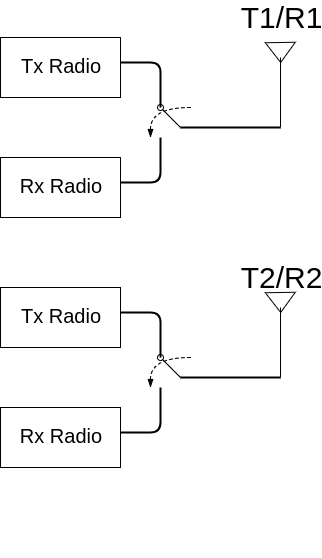}
  \caption{HD node with two transmit and two receive antennas, and four RF chains }
  \label{fig:HD}
\end{subfigure}\\[1ex]
\begin{subfigure}{.45\linewidth}
  \centering
  \includegraphics[scale=0.28]{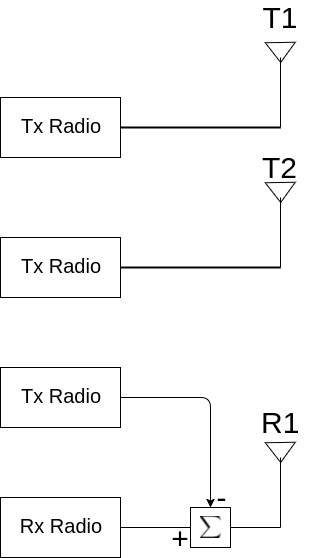}
  \caption{RC FD node with two transmit and one receive antennas, and four RF chains}
  \label{fig:FD_RC}
\end{subfigure}
\begin{subfigure}{.45\linewidth}
  \centering
  \includegraphics[scale=0.28]{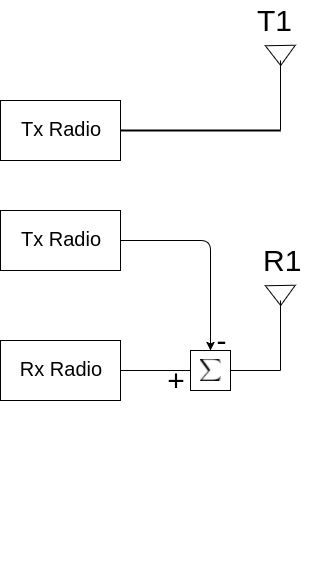}
  \caption{AC FD node with one transmit and one receive antennas, and three RF chains}
  \label{fig:FD_AC}
\end{subfigure}
\caption{Illustration of HD, AC FD and RC FD implementations}
\label{fig:RC_AC}
\end{figure}

For a fair comparison of HD and FD communications, hardware
resources must be equalized. We investigate two conservation
scenarios: antenna conservation, where the number of antennas is
kept equal, and RF chain conservation, where the number of RF chains
is kept equal~\cite{Sanaz}. For instance, if a node has $N$ antennas
in HD mode, then it would have total $2N$ RF chains ($N$ each for
up-converting and down-converting). While considering AC FD, we take
total number of antennas to be $N$, i.e., if $r$ antennas are used
for reception then remaining $(N-r)$ antennas are used for
transmission. Whereas for the RC FD implementation, the total number
of RF chains is kept same as that in HD case, which is $2N$. Hence
if $r$ antennas are used for reception, then in addition to $r$
down-converting RF chains, $r$ RF chains are used in analog
cancellation, and remaining $2N-2r$ RF chains can be used for
up-converting in transmission, resulting in $2N-2r$ transmit
antennas. Note that, RC FD increases the total number of antennas in
the FD mode. This is illustrated in Figure~\ref{fig:RC_AC}, where we consider a HD node with 
two antennas ($N=2$), and  hence it has four RF chains (Figure~\ref{fig:HD}). In RC FD implementation, the total number of RF chains is four, resulting in two transmit and one receive antennas, since one RF chain is required for the analog cancellation (Figure~\ref{fig:FD_RC}). In AC FD scenario there are two antennas, one each for transmission and reception (Figure~\ref{fig:FD_AC}).

Comparison of the number of antennas is summarized in
Table \ref{tab:tab1}. (See also~\cite{Sanaz,Aggarwal12}.)
\begin{table}[h]
\centering \scalebox{1.00}{
\begin{tabular}{ |c|c|c|c| }
 \hline
 & number of  RX  & number of  TX  & Total number of  \\
 & antennas &  antennas & antennas \\
\hline
 HD & N & N & N \\
\hline
AC FD & r & N-r & N\\
\hline
RC FD & r & 2N-2r& 2N-r \\
 \hline
\end{tabular}}
\caption{Number of antennas in HD, AC FD and RC FD implementations}
\label{tab:tab1}
\end{table}

\section{Two Way Channel}
\label{sec:TWC}

As the first scenario, we consider two way channel between nodes $A$
and $B$, as illustrated in Figure \ref{fig:TWC}. In this section, we formulate, calculate and compare the  DoF of two way communication in HD and FD modes.

\subsection{Half-Duplex Mode}
When nodes $A$ and $B$ communicate in HD mode in the same band, they
need to employ time sharing. Hence, the nodes alternate for
transmission, as depicted in Figures \ref{fig:TWC_HD_AB} and
\ref{fig:TWC_HD_BA}. Defining $\tau$ as the fraction of time, in
which node $A$ transmits while node $B$ receives, the remaining
fraction, $\left(1-\tau\right)$ is utilized by node $B$ for transmission while
node $A$ receives.
\subsubsection {Achievable Rate}
Recalling that the SINR at node $B$, $\Gamma_{AB}$ is calculated via
equation (\ref{eq:SNR}), with the SI term $I_B(P_B)$ as zero in HD
mode, the average achievable rate from $A$ to $B$ $R^{HD}_{AB}$ can
be obtained as \cite{Tse}, \beq R^{HD}_{AB}=\tau
\mathbb{E}\left[\log\det\left(\mathbf{I}+\frac{\Gamma_{AB}}{N_A}\mathbf{H}_{AB}\mathbf{H}_{AB}^*\right)\right],
\label{eq:R_AB_HD_TWC} \eeq \beq R^{HD}_{BA}=\left(1-\tau\right)
\mathbb{E}\left[\log\det\left(\mathbf{I}+\frac{\Gamma_{BA}}{N_B}\mathbf{H}_{BA}\mathbf{H}_{BA}^*\right)\right],
\label{eq:R_BA_HD_TWC} \eeq where $\mathbf{I}$ denotes the identity
matrix, with $\mathbf{I} \in \mathcal{C}^{N_{B}\times N_{B} }$ for
(\ref{eq:R_AB_HD_TWC}) and $\mathbf{I} \in \mathcal{C}^{N_{A}\times
N_{A} }$ for (\ref{eq:R_BA_HD_TWC}), and $N_X$ is the total number
of antennas for node $X$.
\subsubsection{Degrees of Freedom}
The DoF characterizing the performance at high SNR, for the two way
channel considering HD communication is obtained as follows:
\begin{align*}
\text{DoF}^{HD}_{AB}&=\lim_{P_A \to \infty}\frac{R_{AB}^{HD}}{\log(P_A)}=\tau \min(N_A,N_B), \nn
\text{DoF}^{HD}_{BA}&=\lim_{P_B \to
\infty}\frac{R_{BA}^{HD}}{\log(P_B)}=(1-\tau) \min(N_A,N_B).
\end{align*}
This results in the following DoF trade-off,
\begin{align}
\left\{\text{DoF}^{HD}_{AB},\text{DoF}^{HD}_{BA}\right\}=\left\{\tau
,(1-\tau)\right\} &\min(N_A,N_B), \nn
 &0\le\tau\le 1. 
 \label{EQ:TWC_HD_TO}
\end{align}

\subsection{Full-Duplex Mode}
In this case, both nodes $A$ and $B$ are assumed to have FD
capability, so that they can transmit to each other simultaneously
in the same band. Then the SINR per node, $\Gamma$ is calculated
from (\ref{eq:SNR}) with the residual SI model in
(\ref{eq:SImodel1}). Below, $t_X$ denotes the number of antennas
used for transmission, and $r_X$ is the number of antennas used for
reception at node $X$ in FD mode, as illustrated in Section
\ref{sec:hardware_resource}.

\subsubsection{Achievable Rates}

The average achievable rates can be calculated as, \beq R_{AB}=
\mathbb{E}\left[\log\det\left(\mathbf{I}+\frac{\Gamma_{AB}}{t_A}\mathbf{H}_{AB}\mathbf{H}_{AB}^*\right)\right],
\label{eq:R_AB_FD_TWC} \eeq \beq
R_{BA}=\mathbb{E}\left[\log\det\left(\mathbf{I}+\frac{\Gamma_{BA}}{t_B}\mathbf{H}_{BA}\mathbf{H}_{BA}^*\right)\right],
\label{eq:R_BA_FD_TWC} \eeq where $\mathbf{I} \in
\mathcal{C}^{r_{B}\times r_{B} }$ for (\ref{eq:R_AB_FD_TWC}) and
$\mathbf{I} \in \mathcal{C}^{r_{A}\times r_{A} }$ for
(\ref{eq:R_BA_FD_TWC}).

\subsubsection{Degrees of Freedom}
\label{sec:TWC_FD_DoF}
The DoF trade-off for FD mode is achieved through the following
power scaling approach,
\begin{equation}
\frac{\log(P_B)}{\log(P_A)}=\gamma,
\end{equation}
for some $\gamma>0$. Thus, the achievable DoF from node A to B are calculated as
\begin{align*}
\text{DoF}^{FD}_{AB} =\lim_{\substack{P_A \to \infty\\ P_B=P_A^\gamma}}\frac{R_{AB}^{FD}}{\log(P_A)}
= [1-\gamma(1-\lambda)]^+\min(r_B, t_A).
\end{align*}
Similarly, from node B to A,
\begin{align*}
\text{DoF}^{FD}_{BA} =\lim_{\substack{P_B \to \infty\\ P_B=P_A^\gamma}}\frac{R_{BA}^{FD}}{\log(P_B)}
= \left[1-\frac{(1-\lambda)}{\gamma}\right]^+\min(r_A, t_B).
\end{align*}
Hence following DoF trade-off region is achievable:
\begin{align}
\left\{\text{DoF}^{FD}_A,\text{DoF}^{FD}_B\right\}=\bigcup_{ \substack{r_A,r_B,\\t_A,t_B}}\bigg\{ [1-\gamma(1-\lambda)]^+\min(r_B,
t_A),\nn
\left[1-\frac{(1-\lambda)}{\gamma}\right]^+\min(r_A,
t_B)\bigg\}. \label{EQ:TWC_FD_DoF}
\end{align}
Here $\bigcup$ denotes the convex-hull over the admissible
parameters. The possible ranges for $r_A, r_B, t_A$ and $t_B$ depend
on the hardware constraints as shown in Table \ref{tab:tab1}.

\subsection{Comparison of the HD and FD Modes}
\label{subsec:comp_twc}
Below, we evaluate and compare the achievable DoF  of two way communication in HD and FD modes, considering
different transmission power levels, number of antennas and SI
cancellation levels. For the FD mode, we refer to the two
implementation models considering the allocation of radio resources,
namely, the AC FD and RC FD, as described in Section \ref{sec:hardware_resource}.

From the DoF perspective, the following
proposition shows that in the AC case, HD performs better than the achievable FD DoF trade-off. 
\newtheorem{prop}{Proposition}
\begin{prop}
When $\lambda<1$, the achievable DoF region for AC FD implementation of the two
way channel lies strictly inside the HD implementation.
\end{prop}
\begin{IEEEproof}
The DoF region in (\ref{EQ:TWC_HD_TO}) is equivalent to the following
\begin{equation*}
\text{DoF}^{HD}_{AB}+\text{DoF}^{HD}_{BA}=\min(N_A,N_B).
\end{equation*}
Hence, in order to show that the FD DoF region lies strictly inside
the HD one for $\lambda<1$, it suffices to show that
\begin{equation*}
\text{DoF}^{FD}_{AB}+\text{DoF}^{FD}_{BA}<\min(N_A,N_B).
\end{equation*}
Note that when we are operating at a point when both DoF's are strictly positive, (\ref{EQ:TWC_FD_DoF}) for the AC scenario becomes
\begin{align}
\text{DoF}^{FD}_{AB}&=(1-\gamma(1-\lambda))\min(t_A, N_B-t_B), \nonumber \\
\text{DoF}^{FD}_{BA}&=\left(1-\frac{1-\lambda}{\gamma}\right)\min(N_A-t_A, t_B),
\label{EQ:FD_TO_AC}
\end{align}
for some $\gamma\in\left[1-\lambda,  (1-\lambda)^{-1}\right]$
and $0<t_A<N_A$, $0<t_B<N_B$. Since $1-\lambda \le\gamma\le
(1-\lambda)^{-1}$, we have
\begin{equation}
1-\gamma(1-\lambda) \le 1-(1-\lambda)^2,
\label{eq:ub1}
\end{equation}
and also
\begin{equation}
1-\frac{1-\lambda}{\gamma}\le 1-(1-\lambda)^2.
\label{eq:ub2}
\end{equation}
Then from (\ref{EQ:FD_TO_AC}), (\ref{eq:ub1}) and (\ref{eq:ub2})
\begin{align}
\text{DoF}^{FD}_{AB}&\le( 1-(1-\lambda)^2)\min(t_A, N_B-t_B), \nonumber \\
\text{DoF}^{FD}_{BA}&\le( 1-(1-\lambda)^2)\min(N_A-t_A, t_B).
\label{EQ:FD_TO_AC_UB}
\end{align}
Thus,
\begin{align*}
\text{DoF}^{FD}_{AB}+\text{DoF}^{FD}_{BA} & \le  ( 1-(1-\lambda)^2)(\min(t_A, N_B-t_B)\\
&\quad+\min(N_A-t_A, t_B)) \\
&\le ( 1-(1-\lambda)^2)\min(N_A,N_B) \\
&<\min(N_A,N_B),
\end{align*}
where the last equation follows, since $\lambda<1$ implies $1-(1-\lambda)^2)<1$.
\end{IEEEproof}

The next proposition shows that, unlike AC FD case, for RC FD
implementation, some part of FD DoF region lies outside the HD
region.
\begin{prop}
When $\lambda>\frac{3}{4}\frac{\min(N_A,N_B)}{\min(N_A,N_B)-6}$,
there exists a point in FD DoF region, which is not achievable by HD
transmission for the RF conserved scenario. If $N_A$ and $N_B$ are
divisible by $3$ then the condition becomes $\lambda>3/4$.
\end{prop}
\begin{IEEEproof}
Considering $\gamma=1$, $r_A=\lfloor2N_A/3\rfloor$ and
$r_B=\lfloor2N_B/3\rfloor$ in (\ref{EQ:TWC_FD_DoF}), results in
$t_A=\lfloor2N_A/3\rfloor$ and $t_B=\lfloor2N_B/3\rfloor$  for the
RC implementation, and
\begin{align}
\text{DoF}^{FD}_A=\lambda\min(\lfloor2N_A/3\rfloor,\lfloor2N_B/3\rfloor),\nn 
\text{DoF}^{FD}_B=\lambda\min(\lfloor2N_A/3\rfloor,\lfloor2N_B/3\rfloor) .
\label{EQ:FD_TO_AC_UB}
\end{align}
Thus,
\begin{align}
\text{DoF}^{FD}_A+\text{DoF}^{FD}_B & = 2\lambda\min(\lfloor 2N_A/3\rfloor,\lfloor 2N_B/3\rfloor) \nonumber \\
&\ge2\lambda\min(2N_A/3-1,2N_B/3-1) \nonumber \\
&=\frac{4}{3}\lambda\min(N_A,N_B)-2\lambda\nonumber \\
&>\min(N_A,N_B)
\end{align}
where the last inequality can be shown to be true after some
algebraic manipulation, when
\begin{align*}
\lambda>\frac{\min(N_A,N_B)}{\frac{4}{3}\min(N_A,N_B)-2}.
\end{align*}
When both $N_A$ and $N_B$ are divisible by 3, there is no flooring
operation, and hence the condition simplifies to $\lambda>3/4$.
\end{IEEEproof}
\begin{figure}[]
\centering
\hspace*{-25mm}
\includegraphics[scale=.30]{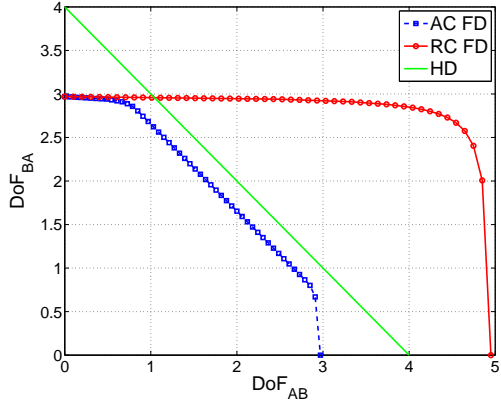}
\caption{Degrees of Freedom region for two way channel,
$N_A=4,N_B=6,\lambda=0.9$}
\label{Fig:TWC_DoF_RC}
\end{figure}
Figure \ref{Fig:TWC_DoF_RC} shows the DoF region for HD, AC FD and RC
FD scenarios, where for the FD case we have plotted the convex hull
in (\ref{EQ:TWC_FD_DoF}). The corner point of the FD trade-off occurs when SI at one of the nodes is so high (due to high transmission power at that node) that the DoF it receives is effectively zero, even though the other node is transmitting to it. Note that while RC FD can achieve DoF pairs not possible with HD, its DoF region does not contain that of
HD. 

Next we argue that additional transmit precoding/receive processing to mitigate SI would not improve the FD DoF found above. In order to do this, we will model a two way channel  as a full-rank 4-user MIMO interference channel, where transmitting nodes of the interference channel correspond to the transmitting units of the nodes $A$ and $B$, and the receiving nodes correspond to the receiving units. Then, SI cancellation can be viewed as  interference cancellation at  each receiving node. We further assume $\lambda=0$ and the SI is Gaussian to be able to carry out the signal analysis. Using the MIMO interference channel results in~\cite{Jafar} we can show  that the maximum sum DoF  for the AC scenario with precoding is $\min(N_A-1, N_B-1)$, obtained with a linear zero forcing precoder. From (\ref{EQ:TWC_FD_DoF}), we see that when $\lambda=0$, substituting $\gamma=0$ (or $\gamma=\infty$) yields the sum DoF of $\min(N_A-1, N_B-1)$. Since for $\lambda>0$, the sum DoF in (\ref{EQ:TWC_FD_DoF}) would be strictly better, we conclude that transmit precoding does not improve the achievable two way channel DoFs found in this paper. Another advantage of the hardware SI mitigation technique adopted in this paper is that accurate estimates of the SI matrix are not required to implement precoding in the signal space.  
\section{Two Hop Channel: Relaying}
\label{sec:OWTHC}
As the second scenario, we consider the two hop
channel, where $A$ communicates with $B$ through $R$, as illustrated
in Figure \ref{fig:OWRC}. Assuming that  $B$ does not hear  $A$, we
formulate, calculate and compare the  DoF of two
hop communication, i.e., relaying, in HD and FD modes.
\subsection{Half-Duplex Mode}
\label{sec:HD} In the HD mode, $A$ first transmits to $R$ for a
fraction of time, $\tau$, $R$ decodes the received bits and forwards
them to $B$ for the remaining fraction, $1-\tau$ of time.
\subsubsection{Achievable Rate}
The average rate achievable from $A$ to $R$ is calculated as
\beq
R_{AR}^{HD}=\tau\mathbb{E}\left[\log \det
\left(\mathbf{I}+\frac{\Gamma_{AR}}{
N_A}\mathbf{H}_{AR}\mathbf{H}_{AR}^*\right)\right], \nonumber
\eeq
and the
rate achievable over $R$ to $B$ is given by
\beq
R_{RB}^{HD}=(1-\tau)\mathbb{E}\left[\log \det
\left(\mathbf{I}+\frac{\Gamma_{RB}}{N_R}\mathbf{H}_{RB}\mathbf{H}_{RB}^*\right)\right].
\nonumber
\eeq
By optimizing over $\tau$, the end-to-end average achievable rate
for HD relaying can be found as \beq R_{HD}^{AB}=\max_{0\le\tau \le
1}\min\left(R_{AR}^{HD},R_{RB}^{HD}\right).
\label{eq:R_SD_HD}
\eeq
\subsubsection{Degrees of Freedom}
For the DoF of the two hop channel, as in Section \ref{sec:TWC_FD_DoF} we assume that the relay
scales its power with respect to the transmission power of node A,
according to
\beq
\frac{\log(P_R)}{\log(P_A)}=\gamma, \,\,\,  \, 0<\gamma\le1. \nonumber
\eeq Then, the  DoF
of the relay network in HD mode is given by
\begin{align}
\text{DoF}_{HD}=\sup_{0<\gamma\le1}\lim_{\substack{P_A\to \infty  \\ P_R=P_A^\gamma}}&\frac{R_{HD}}{\log(P_A)} \\
=\max_{\substack{ 0<\tau<1 \\ 0<\gamma\le 1}}\min (&\tau\min(N_A,N_R),\nonumber\\
&\gamma(1-\tau)\min(N_R,N_B)) \label{eq:C}\\
= \max_{0<\tau<1} \min &(\tau N_A,\tau N_R,\nonumber\\
&(1-\tau)N_R,(1-\tau
)N_B).
\end{align}
Note that, in (\ref{eq:C}), setting $\gamma=1$ maximizes the
$\text{DoF}_{HD}$. Depending on the values of $N_A$, $N_R$, and
$N_B$, and using optimal $\tau$ denoted as $\tau_{opt}$, we obtain
following DoF values for the HD mode as shown in the Table
\ref{tab:DOFHD}.

\begin{table}[H]
\centering
\begin{tabular}{| c | c | c |}

\hline
   & $\tau_{opt}$ & $\text{DoF}_{HD}$ \\ [8pt] \hline
  $N_R \le \min({N_A,N_B})$ & $\frac{1}{2}$ & $\frac{N_R}{2}$ \\ [8pt] \hline
  $N_R \ge \max({N_A,N_B})$ & $\frac{N_B}{N_B+N_A}$ & $\frac{N_AN_B}{N_B+N_A}$ \\ [8pt] \hline
  $N_A \le N_R \le N_B$ & $\frac{N_R}{N_R+N_A}$ & $\frac{N_RN_A}{N_R+N_A}$ \\ [8pt]\hline
  $N_B \le N_R \le N_A$ & $\frac{N_B}{N_B+N_R}$ & $\frac{N_RN_B}{N_R+N_B}$ \\ [8pt] \hline

\end{tabular}
\caption{Degrees of Freedom for HD relaying}
\label{tab:DOFHD}
\end{table}
\subsection{Full-Duplex Mode}
\label{sec:OWRC_FD} In FD relaying, $R$ is able to receive and
transmit simultaneously in the same band, however, it is subject to
SI. In order to maintain causality, the relay node transmits
$(i-1)^{\text{th}}$ symbol, while it receives the $i^{\text{th}}$
symbol.
\subsubsection{Achievable Rate}
When the relay node operates in FD mode, the rates are calculated as
follows:
\begin{align}
R_{AR}^{FD}=\mathbb{E}\left[\log \det \left(\mathbf{I}+\frac{\Gamma_{AR}}{N_A}\mathbf{H}_{AR}\mathbf{H}_{AR}^\ast\right)\right],
\label{eq:R_AR_FD}\nn
R_{RB}^{FD}=\mathbb{E}\left[\log \det \left(\mathbf{I}+\frac{\Gamma_{RB}}{t}\mathbf{H}_{RB}\mathbf{H}_{RB}^\ast\right)\right].
\end{align}
Recall that $t=(N_R-r)$ for AC FD, and $t=(2N_R-2r)$ for RC FD.
Depending on the average SINR at the relay node and SNR at $B$, the
excess power at the relay can have a negative impact on the
achievable rate due to increased SI. In fact, the SINR at the relay
node is decreased as the relay power $P_R$ is increased,
while $P_A$ is held constant. Thus, with the increase in
$P_R$ for a constant $P_A$, the rate of the channel from 
node $A$ to $R$ is decreased, while the rate of the
channel from $R$ to node $B$ is increased. Therefore, by letting ${P_R}_{max}$ denote the maximum average power at the relay, the achievable rate for FD relaying can be written as
\begin{equation}
R_{FD}^{AD}=\max_{ \substack{ 0<r<N_R \\   P_R \le {P_R}_{max}}}\min\left(R_{AR}^{FD},R_{RB}^{FD}\right).
\label{eq:R_SD_FD_1}
\end{equation}
\subsubsection{Degrees of Freedom}
Assuming power scaling as in (\ref{eq:C}), the DoF of FD relaying is obtained as
\beq
\text{DoF}^{FD}_{AB}=\sup_{0<\gamma\le1}\lim_{\substack{{P}_A \to \infty
\\ {P}_R =P_A^\gamma}}\frac{R_{FD}}{\log(P_A)}, \nonumber
\eeq
where in order to control the SI, the relay  scales its power with respect to the transmission power of node $A$, through
 \beq
 \frac{\log(P_R)}{\log(P_A)}=\gamma, \,\,\,  \, 0<\gamma\le1. \nonumber
 \eeq
Then, the achievable DoF for FD relaying can be computed as
\begin{align}
\text{DoF}^{FD}_{AB}=\max_{\substack{0<r<N_R \\ 0<\gamma\le 1}}\min (&(1-\gamma(1-\lambda))\min(N_A,r),\nonumber\\
&\gamma\min(t,N_B)) \nonumber\\
=\max_{\substack{0<r<N_R \\ 0<\gamma\le 1}}\min (&(1-\gamma(1-\lambda))N_A,(1-\gamma(1-\lambda))r,\nonumber\\
&\gamma t,\gamma N_B).
\label{eq:OWRC_DoF_FD}
\end{align}
Here, $t=(N_R-r)$ for the AC FD implementation and $t=(2N_R-2r)$ for
the RC FD implementation. We explicitly compute the
$\text{DoF}_{AB}^{FD}$ for symmetric and asymmetric cases and
compare it with $\text{DoF}_{AB}^{HD}$ in the next subsection.

\subsection{Comparison of HD and FD Relaying}
\label{sec:Relay_comp}
\begin{itemize}
\item \textbf{Symmetric Case ($N_A=N_B$)}:\\
To compare the DoF of HD relaying and FD relaying, we first consider
a symmetric case when the $A$ and $B$ have same number of antennas,
i.e., $N_A$=$N_B$=$N$, and $N_R$ is even. Using
the Table \ref{tab:DOFHD}, one can obtain
\begin{equation}
\text{DoF}^{HD}_{AB}=\min\left(\frac{N}{2},\frac{N_R}{2}\right).
\label{EQ:DoF_OWRC_HD}
\end{equation}

For AC FD implementation
(\ref{eq:OWRC_DoF_FD}) can be written as
\begin{align}
\text{DoF}^{FD,AC}_{AB}&=\max_{\substack{1\le r < N_R \\  0<\gamma\le 1}}\min ((1-\gamma(1-\lambda))N,\nonumber\\
&\qquad\qquad(1-\gamma(1-\lambda))r, (N_R-r),\gamma N) \nonumber  \\
&\le \max_{0<\gamma\le 1}\min ((1-\gamma(1-\lambda))N,\gamma N) \label{eq:DoF1}\\
&=\frac{N}{2-\lambda}. \label{eq:DoF2}
\end{align}
Since the minimum of the two terms in (\ref{eq:DoF1}) is maximized
when both are equal, (\ref{eq:DoF2}) can be obtained by setting
$\gamma=\frac{1}{2-\lambda}$. Similarly,
\begin{align}
\text{DoF}^{FD,AC}_{AB} &\le \max_{\substack{1\le r < N_R \\ 0<\gamma\le 1}}\min ((1-\gamma(1-\lambda))r,\gamma(N_R-r)),\nonumber\\
&=\frac{N_R}{2(2-\lambda)} \label{eq:DoF3},
\end{align}
where (\ref{eq:DoF3}) is obtained by setting $\gamma=\frac{1}{2-\lambda}$ and $r=\frac{N_R}{2}$.

From (\ref{eq:DoF2}) and  (\ref{eq:DoF3}), we can write
\begin{equation}
\text{DoF}^{FD,AC}_{AB} \le \frac{1}{2-\lambda}\min\left(N,\frac{N_R}{2}\right).
\label{eq:DoF4}
\end{equation}
Equality in (\ref{eq:DoF4}) can be achieved when $\gamma=\frac{1}{2-\lambda}$ and $r=\frac{N_R}{2}$, leading to
\begin{equation}
\text{DoF}^{FD,AC}_{AB} =\frac{1}{2-\lambda}\min\left(N,\frac{N_R}{2}\right).
\label{EQ:DoF_OWRC_FD_AC}
\end{equation}

Similarly, for the RC FD implementation we have,
\begin{equation}
\text{DoF}^{FD,RC}_{AB}=\frac{1}{2-\lambda}\min\left(N,\left\lfloor\frac{2N_R}{3}\right\rfloor\right).
\label{EQ:DoF_OWRC_FD_RC}
\end{equation}

\begin{figure}[]
\hspace*{-25mm}
\includegraphics[scale=0.30]{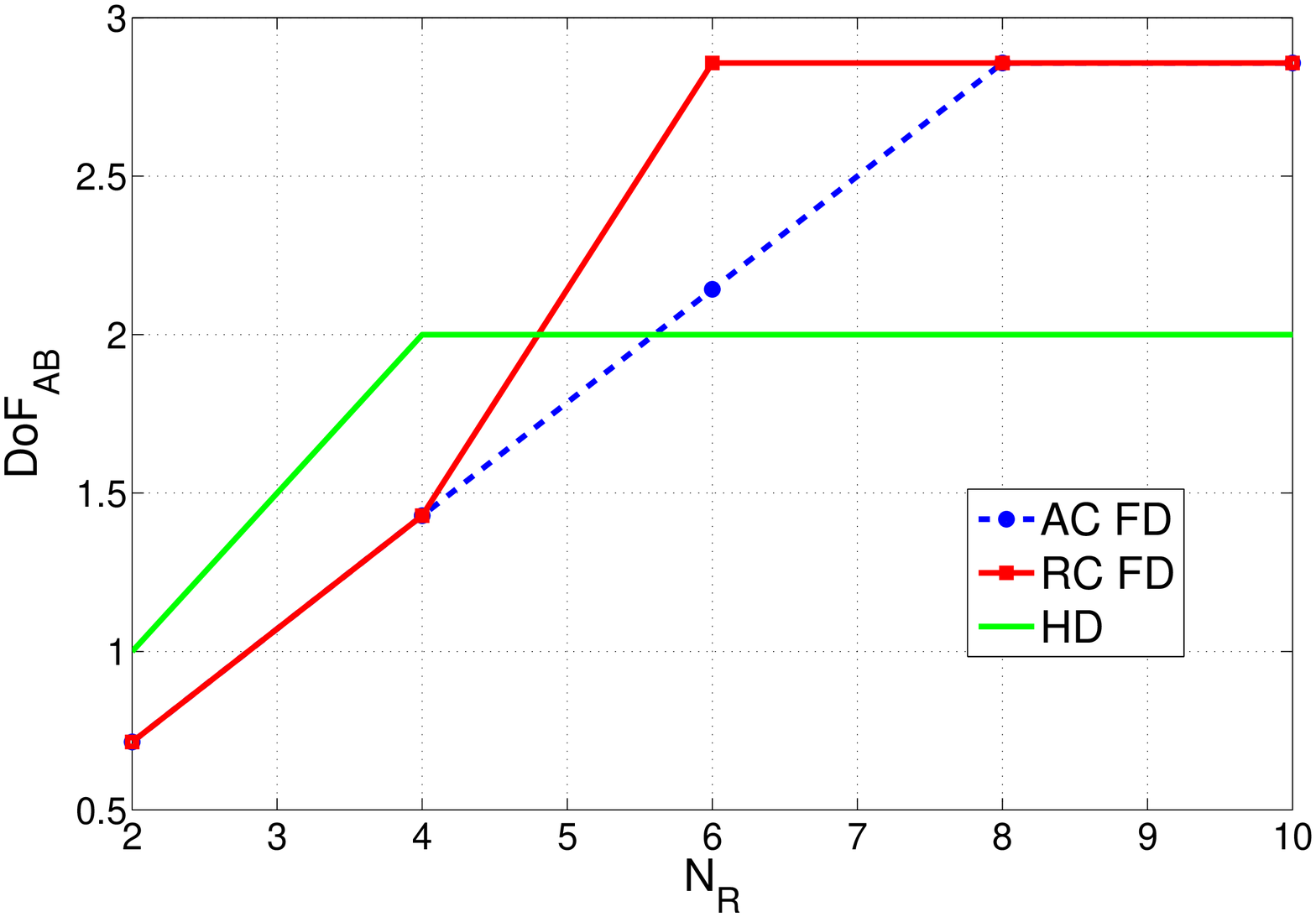}
\caption{DoF for HD relaying and FD relaying, $N_A$=$N_B$=$4$.} \label{fig:DoF_THC}
\end{figure}

The DoF results for this case are plotted in Figure
\ref{fig:DoF_THC}. As seen from the figure, when $N_R$ is small, HD
relaying performs better than FD relaying, and the situation is
reversed when $N_R$ gets larger, with the RC FD implementation
always dominating AC FD implementation. Comparing
(\ref{EQ:DoF_OWRC_HD}) and (\ref{EQ:DoF_OWRC_FD_AC}),
$\text{DoF}^{FD,AC}_{AB}>\text{DoF}^{HD}_{AB}$ if
\begin{align}
N_R>N(2-\lambda) \quad \text{and} \quad \lambda>0.
\label{EQ:DoF_OWRC_comp_AC}
\end{align}
Similarly from (\ref{EQ:DoF_OWRC_HD}) and (\ref{EQ:DoF_OWRC_FD_RC}), if $N_R$ is divisible by $3$, then $\text{DoF}^{FD,RC}_{AB}>\text{DoF}^{HD}_{AB}$ provided
\begin{align}
N_R>\frac{3}{4} N(2-\lambda) \quad \text{and} \quad \lambda>0.
\label{EQ:DoF_OWRC_comp_RC}
\end{align}
If $N_R$ is not divisible by $3$, then (\ref{EQ:DoF_OWRC_HD}) and (\ref{EQ:DoF_OWRC_FD_RC}) can be evaluated to compare  $\text{DoF}^{FD,RC}_{AB}$ and $\text{DoF}^{HD}_{AB}$.

\item \textbf{Asymmetric Case ($N_A=1$)}:\\
Now, we consider the asymmetric case, where node $A$ has a single
antenna, i.e., $N_A=1$, whereas the relay  and node $B$ have
multiple antennas, $N_R$ and $N_B\ge 1$, respectively. This could model a
cellular phone, which cannot afford multiple antennas, communicating
with an access point through a relay.

From the general expression for the DoF for the FD relaying channel for both AC and RC implementation,
(\ref{eq:OWRC_DoF_FD}) with $N_A=1$
\begin{align}
\text{DoF}^{FD}_{AB}
&=\max_{\substack{0<r<N_R \\ 0<\gamma \le 1}}\min ((1-\gamma(1-\lambda)),(1-\gamma(1-\lambda))r,\nonumber\\
&\quad\qquad\qquad\qquad \gamma (N_R-r),\gamma N_B)\label{EQ:OWRC_DoF_FD_r1}\\
&=\max_{ 0<\gamma\le 1}\min ((1-\gamma(1-\lambda)),\gamma(N_R-1),\gamma N_B)\label{EQ:OWRC_DoF_FD_r2}\\
&=\frac{\min(N_R-1,N_B)}{\min(N_R-1,N_B)+1-\lambda}
\end{align}
where we have set $r=1$ in (\ref{EQ:OWRC_DoF_FD_r2}) since the minimum of first two terms in (\ref{EQ:OWRC_DoF_FD_r1}) does not depend on $r$ and the third term is decreasing in $r$.

The DoF for HD relaying yields
\begin{equation*}
\text{DoF}^{HD}_{AB}=\frac{\min(N_R,N_B)}{\min(N_R,N_B)+1},
\end{equation*}
It can be seen that $\text{DoF}^{FD}_{AB}>\text{DoF}^{HD}_{AB}$ if
\begin{equation*}
\lambda > 1-\frac{\min(N_R-1,N_B)}{\min(N_R,N_B)}.
\end{equation*}

Hence, for both AC FD and RC FD implementations, the following
holds: If $N_R>N_B$, then DoF of FD relaying is strictly larger than
DoF of HD relaying for both AC FD and RC FD implementations,
provided $\lambda>0$. If $N_R\le N_B$, then FD relaying performs
better than HD relaying if $\lambda>\frac{1}{N_R}$. Thus the FD implementation is better than the HD  if
\begin{align}
N_R>\min\left(N_B,\frac{1}{\lambda}\right).
\label{EQ:OWRC_FD_DOF_Assym_Comp}
\end{align}
\end{itemize}

It is interesting to note that, for the case $N_A=N_B=1$, and
$N_R=2$, \cite{Rodriguez2} obtained the DoF (referred to as
multiplexing gain in \cite{Rodriguez2}) for FD relaying channel, with the
relay node operating in amplify and forward mode, using a similar SI
model as in this paper. Their multiplexing gain term of
$\frac{1}{2-\lambda}$ matches with our DoF, when evaluated at $N=1$,
and $N_R=2$ in (\ref{EQ:DoF_OWRC_FD_AC}).

\section{Two Way Two Hop Channel: Two Way Relaying }
\label{sec:TWTHC} As the third scenario, we consider the two way two
hop channel, where $A$ and $B$ communicate with each other through
$R$, performing two way relaying, as illustrated in Figures \ref{fig:TWRC1} and \ref{fig:TWRC2}
representing HD and FD modes, respectively. Again, there is no
direct link between nodes $A$ and $B$. We formulate, calculate and
compare the achievable rates as well as DoF of two way two hop
communication in HD and FD modes.

\subsection{Half-Duplex Mode}
\subsubsection{Achievable Rates}
In HD mode, with transmission strategies described in the
corresponding system model in Section \ref{system_models}
, the average achievable
rates can be calculated as follows: During the MAC phase, both nodes
$A$ and $B$ transmit their messages to the relay node with
achievable rates calculated as \cite{Tse},
\begin{gather*}
R^{HD}_{AR} \leq \tau\mathbb{E}\left[\log\det\left(\mathbf{I}+\frac{{ \Gamma }_{AR}}{ N_A}\mathcal{H}_{AR}\mathbf{H}_{AR}^\ast\right)\right],\nn
R^{HD}_{BR} \leq \tau\mathbb{E}\left[\log\det\left(\mathbf{I}+\frac{ \Gamma _{BR}}{ N_B}\mathbf{H}_{BR}\mathbf{H}_{BR}^\ast\right)\right],\nonumber\\
R^{HD}_{AR}+R^{HD}_{BR}\leq\tau\mathbb{E}\left[\log\det\left(\mathbf{I}+\frac{{ \Gamma }_{AR}}{ N_A}\mathbf{H}_{AR}\mathbf{H}_{AR}^\ast\right.\right.\\
\qquad\qquad\qquad\qquad+\left.\left.\frac{ \Gamma _{BR}}{ N_B}\mathbf{H}_{BR}\mathbf{H}_{BR}^\ast\right)\right].
\end{gather*}
Here we assume that MAC phase lasts for $\tau$ fraction of the time. During the BC phase, the relay node broadcasts a message to both of the nodes, such that each node can retrieve the other node's message
by subtracting its own data. The achievable rates for this phase are
obtained as \cite{Tse},
\begin{align*}
R^{HD}_{RA} \leq \left(1-\tau\right)\mathbb{E}\left[\log\det\left(\mathbf{I}+\frac{{ \Gamma }_{RA}}{ N_R}\mathbf{H}_{RA}\mathbf{H}_{RA}^\ast\right)\right],\\
R^{HD}_{RB} \leq \left(1-\tau\right)\mathbb{E}\left[\log\det\left(\mathbf{I}+\frac{{ \Gamma }_{RB}}{ N_R}\mathbf{H}_{RB}\mathbf{H}_{RB}^\ast\right)\right].
\end{align*}
Then, the end-to-end rates are obtained as
\begin{align*}
R^{HD}_{AB} =\min \left(R_{AR},R_{RB}\right), 
R^{HD}_{BA} = \min\left(R_{BR},R_{RA}\right).
\end{align*}
The BC phase is assumed to last $(1-\tau)$ fraction of the time.

Note that, dropping the sum rate constraint from the MAC phase gives
the cut-set upper bound for the HD two way two hop channel
\cite{knopp}.

\subsubsection{Degrees of Freedom}
We compute an upper bound for DoF for HD two way two hop
channel, which we later compare with the performance of FD two way two hop
channel. We also compare FD with the achievable MAC-BC HD scheme  introduced above.

During the first phase, which is assumed to be used for the
fraction $\tau$ of time, the $\text{DoF}_{{AR}}$ and
$\text{DoF}_{{BR}}$ are upper bounded by the respective
point-to-point DoFs, i.e.,
\begin{align*}
\text{DoF}_{{AR}}\le\tau\min (N_A,N_R), \, 
\text{DoF}_{{BR}}\le\tau\min (N_B,N_R).
\end{align*}
Similarly, during the second phase, the DoF expressions are
\begin{align*}
\text{DoF}_{RA}\le(1-\tau)\min (N_A,N_R), \\
\text{DoF}_{{RB}}\le(1-\tau)\min (N_B,N_R).
\end{align*}
The achievable DoF with MAC-BC scheme has an additional sum
constraint
\begin{align*}
\text{DoF}_{RA}+\text{DoF}_{{RB}}\le(1-\tau)\min (2N_A,N_R).
\end{align*}
Hence, the upper bound on end-to-end DoF is
\begin{align*}
\text{DoF}_{ {AB}}&\le\min(\text{DoF}_{ {AR}},\text{DoF}_{ {RB}})\nn
&=\min(\tau N_A, \tau N_R, (1-\tau)N_R,(1-\tau)N_B), \nonumber \\
\text{DoF}_{ {BA}}&\le\min(\text{DoF}_{ {BR}},\text{DoF}_{ {RA}})\\
&=\min(\tau N_B, \tau N_R, (1-\tau)N_R,(1-\tau)N_A).
\end{align*}
For the symmetric case $N_A=N_B=N$, $\tau=\frac{1}{2}$ maximally
enlarges the outer bound region, i.e.,
\begin{align*}
\mathtt{DoF}^{HD}_{UB} = \bigg\{&\text{DoF}_{ {AB}},\text{DoF}_{ {BA}} \in \mathbb{R}^{2} :  \\
& \text{DoF}_{ {AB}} \le \min\left(\frac{N}{2},\frac{N_R}{2}\right)  \nonumber\\
  &\text{DoF}_{ {BA}}\le \min\left(\frac{N}{2},\frac{N_R}{2}\right) \bigg\}.
\end{align*}
Similarly, taking $\tau=\frac{1}{2}$ gives the following achievable
DoF region with MAC-BC scheme
\begin{align}
\mathtt{DoF}^{HD}_{DF}=\bigg\{ &\left(\text{DoF}_{ {AB}},\text{DoF}_{ {BA}}\right) \in \mathbb{R}^{2} :\nn
 & \text{DoF}_{ {AB}} \le \min\left(\frac{N}{2},\frac{N_R}{2}\right)  \nonumber\\
  & \text{DoF}_{ {BA}}\le \min\left(\frac{N}{2},\frac{N_R}{2}\right) \nonumber\\
  & \text{DoF}_{ {AB}}+\text{DoF}_{ {BA}}\le \min\left(N,\frac{N_R}{2}\right)\bigg\}.
  \label{EQ:TWRC_HD_DoF_Ach}
\end{align}
\subsection{Full-Duplex Mode}
In this section, we assume only $R$ is FD enabled, and $A$ and $B$
are HD nodes. As described in the corresponding system model in
Section \ref{system_models}, FD two way two hop communication takes
place in two phases assigned for each direction, where $A$ and $B$
send data to each other, as $R$ performs FD relaying. Again, it is
assumed that the fraction of time devoted to first phase is denoted
by $\tau$, and the remaining fraction, $1-\tau$, is assigned to the
second phase.

\subsubsection{Achievable Rates}
According to SINR expressions at the nodes, the average rates are
calculated through the following expressions:
\begin{align*}
R^{FD}_{AR} &\leq \mathbb{E}\left[\log\det\left({\mathbf{I}}+\frac{{ \Gamma }_{AR}}{N_A} \mathbf{H}_{AR} \mathbf{H}_{AR}^\ast\right)\right], \\
R^{FD}_{RB} &\leq \mathbb{E}\left[\log\det\left( {\mathbf{I}}+\frac{{ \Gamma }_{RB}}{t} \mathbf{H}_{RB} \mathbf{H}_{RB}^\ast\right)\right], \nn
R^{FD}_{BR} &\leq \mathbb{E}\left[\log\det\left( {\mathbf{I}}+\frac{{ \Gamma }_{BR}}{N_B} \mathbf{H}_{BR} \mathbf{H}_{BR}^\ast\right)\right], \\
R^{FD}_{RA} &\leq \mathbb{E}\left[\log\det\left( {\mathbf{I}}+\frac{{ \Gamma }_{RA}}{t} \mathbf{H}_{RA} \mathbf{H}_{RA}^\ast\right)\right].
\end{align*}
The end-to-end average rate is given by the following expressions,
\begin{align*}
R^{FD}_{AB} = \tau\min\left(R_{AR},R_{RB}\right),  R^{FD}_{BA}=
(1-\tau)\min\left(R_{BR},R_{RA}\right).
\end{align*}
\subsubsection{Degrees of Freedom}
Since only $R$ is FD capable, the DoF achievable from node $A$ to
node $B$ is given as in Section \ref{sec:OWRC_FD}. Letting
$D^{FD}_{AB}$ and $D^{FD}_{BA}$ be DoFs obtained when communication
takes place from the node $A$ to the node $B$, and in the reverse
direction respectively, the DoF trade off region is obtained via
time sharing as follows:
\begin{align*}
{\mathtt{DoF}}^{FD}=\left\{ \right. & (\text{DoF}^{FD}_{ {AB}},\text{DoF}_{ {BA}}^{FD}) \in \mathbb{R}^{2} : \nn
& \text{DoF}_{ {AB}} \le   \tau D_{AB}^{FD},  
   \text{DoF}_{ {BA}}\le (1-\tau)D_{BA}^{FD} \nonumber\\
 &\left. 0 \le \tau \le 1  \right\}.
\end{align*}
Here, $D_{AB}^{FD}$ is given by (\ref{eq:OWRC_DoF_FD}),
\begin{align*}
D^{FD}_{AB} =\max_{\substack{0<r<N_R \\ 0<\gamma\le 1}}\min (&(1-\gamma(1-\lambda))N_A,(1-\gamma(1-\lambda))r, \\
&\gamma t,\gamma N_B).
\end{align*}
A similar expression holds for $D_{BA}^{FD}$. We explicitly evaluate
and compare the DoF regions of HD and FD two way relaying for some
specific cases in the next subsection.

\subsection{Comparison of HD and FD Two Way Relaying}
Again, we consider the symmetric and
asymmetric scenarios as follows.
\begin{itemize}
\item \textbf{Symmetric Case} ($N_A=N_B$): \\
If the nodes $A$ and $B$ have same number of antennas ($N_A=N_B=N$), and $N_R$ is even
then the DoF region for the AC FD implementation is obtained as from Section~\ref{sec:Relay_comp}
\begin{align*}
{\mathtt{DoF}}^{FD}=\bigg\{  &(\text{DoF}_{ {AB}},\text{DoF}_{ {BA}}) \in \mathbb{R}^{2} : \\
& \text{DoF}_{ {AB}} \le    \frac{\tau}{2-\lambda}\min\left(\frac{N_R}{2},N\right)  \nonumber\\
  & \text{DoF}_{ {BA}}\le \frac{1-\tau}{2-\lambda}\min\left(\frac{N_R}{2},N\right) \nonumber\\
 & 0 \le \tau \le 1  \bigg\}.
\end{align*}
Note that, the corner point of this trade-off is
$\left(\frac{1}{2-\lambda}\min\left(\frac{N_R}{2},N\right),0\right)$,
which is better than the corner point of for the HD case
$\left(\min\left(\frac{N_R}{2},\frac{N}{2}\right),0\right)$ provided 
\begin{equation*}
N_R > (2-\lambda)N \,\text{and}\, \lambda>0.
\end{equation*}
Hence, if $R$ has sufficient number of antennas, then some part of
the FD DoF region lies outside the HD DoF upper bound.

Similarly, for the RC FD implementation, the DoF region is given by
\begin{align}
{\mathtt{DoF}}^{FD}=\bigg\{ & (\text{DoF}_{ {AB}},\text{DoF}_{ {BA}}) \in \mathbb{R}^{2} : \nn
& \text{DoF}_{ {AB}} \le    \frac{\tau}{2-\lambda}\min\left(\left\lfloor\frac{2N_R}{3}\right\rfloor,N\right)  \nonumber\\
  & \text{DoF}_{ {BA}}\le \frac{1-\tau}{2-\lambda}\min\left(\left\lfloor\frac{2N_R}{3}\right\rfloor,N\right) \nonumber\\
 & 0 \le \tau \le 1  \bigg\}.
 \label{EQ:TWRC_FD_DoF_Ach}
\end{align}

Hence the condition for some part of the FD DoF region for RC FD
implementation to lie outside the upper bound of HD one is (provided $N_R$ is divisible by $3$; see (\ref{EQ:DoF_OWRC_comp_RC}))
\begin{equation}
N_R > \frac{3(2-\lambda)}{4}N \,\text{and}\, \lambda>0.
\end{equation}
An example  comparing the DoF trade off for the symmetric case
can be seen in Figure \ref{fig:DoF_TWTHC}, where we have plotted the upper bound for the HD trade-off, an achievable HD trade-off through MAC-BC scheme, and the FD AC and RC trade-off. It can be observed that near the corner points, where one of the node's DoF is small, the AC trade-off is better than the HD trade-off. However near the central region when both of the node's DoF is nearly equal HD trade-off is better.

\begin{figure}[]

 \includegraphics[trim={10cm 0 10cm 0},clip,scale=0.30]{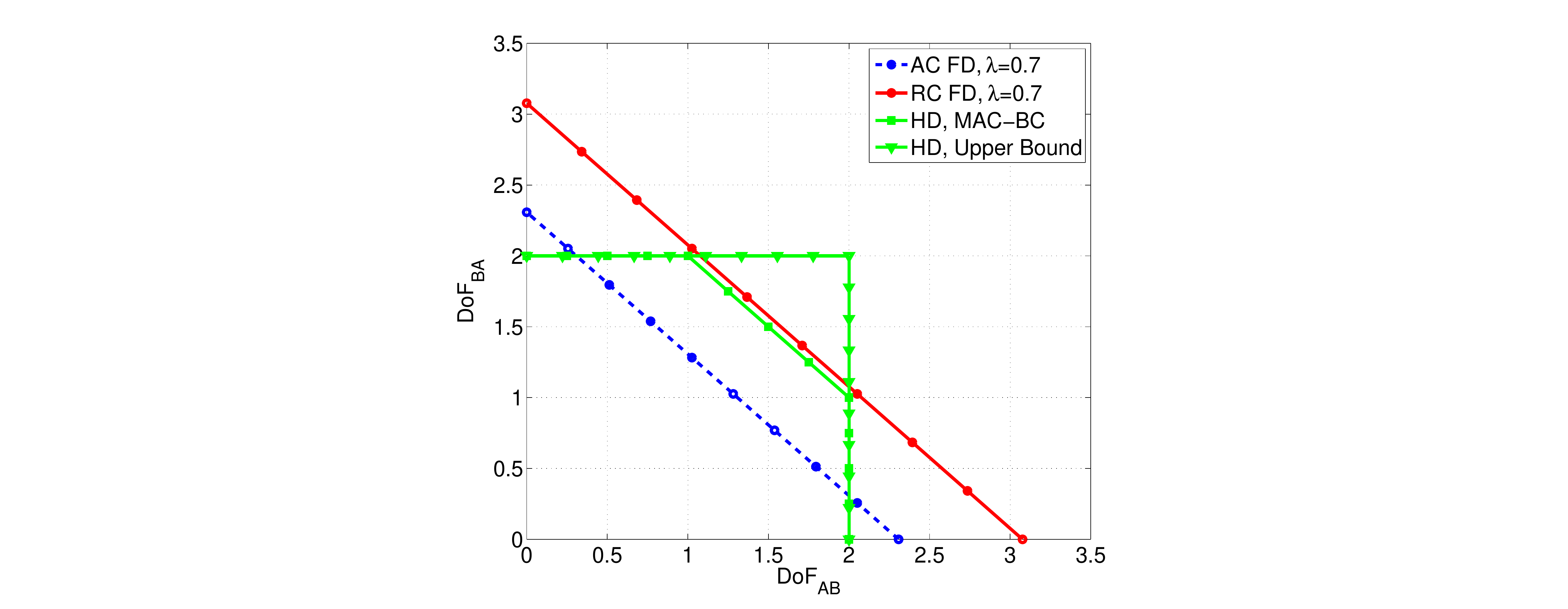}
\caption{DoF trade-off for HD and FD two way relaying,
$N_A=N_B=N=4$, $N_R=6$.} \label{fig:DoF_TWTHC}
\end{figure}

\item \textbf{Asymmetric Case} ($N_A=1$):\\
Using the time sharing and expressions obtained for DoF for the
asymmetric case in Section~\ref{sec:Relay_comp}, the DoF region
for  FD (both AC and RC) can be written as,
\begin{align*}
{\mathtt{DoF}}^{FD}=\Big\{  &(\text{DoF}_{ {AB}},\text{DoF}_{ {BA}}) \in \mathbb{R}^{2} : \\
& \text{DoF}_{ {AB}} \le    \frac{\tau\min(N_R-1,N_B)}{\min(N_R-1,N_B)+1-\lambda}  \nonumber\\
  & \text{DoF}_{ {BA}}\le \frac{(1-\tau)\min(N_R-1,N_B)}{\min(N_R-1,N_B)+1-\lambda} \nonumber\\
 & 0 \le \tau \le 1  \Big\}.
\end{align*}
Comparing with the DoF upper bound for HD, we conclude that some part of the FD DoF region lies
outside the HD one provided (see (\ref{EQ:OWRC_FD_DOF_Assym_Comp}))
\begin{align*}
N_R > \min\left(N_B,\frac{1}{\lambda}\right).
\end{align*}
\end{itemize}

\section{Conclusions}
\label{sec:conclusions}

In this paper, we have compared DoF  for three communication scenarios: two way, two hop, and two way two hop. Using the antenna conserved and RF chain conserved implementations of FD with a realistic residual SI model, we have investigated the conditions under which FD can provide higher throughput than HD. Through detailed DoF analysis, for the two way channel, we have found that the achievable DoF for AC FD is not better than HD with imperfect SI cancellation. For the RC FD case, however, FD DoF trade-off can be better, when the SI cancellation parameter $\lambda$ is high enough. The cross over point depends on various system parameters. In case of the two hop channel, FD is better when the relay has sufficient number of antennas and
$\lambda$ is high enough. For the two way two hop channel, when both
of nodes require similar throughput, the HD implementation is
generally better than FD. However, when one of the terminal's data
rate requirement is significantly higher than the other's (e.g.,
when data flow occurs mostly in one direction, and the other
direction is only used for feedback and control information, or in the case of asymmetric uplink and downlink data rates), then
FD can achieve better DoF pairs than HD, provided the relay has
sufficient number of antennas and the SI suppression factor
$\lambda$ is high enough. 

It should be mentioned that although the DoF results presented for FD are achievable, and the converse results appear to be difficult to obtain, we believe that sophisticated techniques such as zero-forcing, beamforming, or receive processing cannot improve this DoF. Hence the cases for which HD performs better than the achievable FD considered here should still hold true.

The presented results in this paper provide guidelines for choosing HD or FD implementation in practical systems. Future research directions include studying more complex communication scenarios with inter-node
interference and different relaying protocols, where the model and techniques used in this paper may provide a useful foundation.


\bibliographystyle{IEEEtran}
\bibliography{IEEEabrv,HD_FD_f}

\begin{thebibliography}{10}
\providecommand{\url}[1]{#1}
\csname url@samestyle\endcsname
\providecommand{\newblock}{\relax}
\providecommand{\bibinfo}[2]{#2}
\providecommand{\BIBentrySTDinterwordspacing}{\spaceskip=0pt\relax}
\providecommand{\BIBentryALTinterwordstretchfactor}{4}
\providecommand{\BIBentryALTinterwordspacing}{\spaceskip=\fontdimen2\font plus
\BIBentryALTinterwordstretchfactor\fontdimen3\font minus
  \fontdimen4\font\relax}
\providecommand{\BIBforeignlanguage}[2]{{%
\expandafter\ifx\csname l@#1\endcsname\relax
\typeout{** WARNING: IEEEtran.bst: No hyphenation pattern has been}%
\typeout{** loaded for the language `#1'. Using the pattern for}%
\typeout{** the default language instead.}%
\else
\language=\csname l@#1\endcsname
\fi
#2}}
\providecommand{\BIBdecl}{\relax}
\BIBdecl

\bibitem{CISS}
N.~Shende, O.~Gurbuz, and E.~Erkip, ``Half-duplex or full-duplex relaying: A
  capacity analysis under self-interference,'' in \emph{47th Annual Conference
  on Information Sciences and Systems}, March 2013, pp. 1--6.

\bibitem{Duarte10}
M.~Duarte and A.~Sabharwal, ``Full-duplex wireless communications using
  off-the-shelf radios: Feasibility and first results,'' in \emph{Forty Fourth
  Asilomar Conference on Signals, Systems and Computers}, Nov. 2010, pp. 1558
  --1562.

\bibitem{Knox12}
M.~Knox, ``Single antenna full duplex communications using a common carrier,''
  in \emph{13th Annual Wireless and Microwave Technology Conference}, Apr 2012.

\bibitem{choi}
J.~I. Choi, M.~Jain, K.~Srinivasan, P.~Levis, and S.~Katti, ``Achieving single
  channel, full duplex wireless communication,'' in \emph{Proceedings of the
  Sixteenth Annual International Conference on Mobile Computing and
  Networking}, 2010, pp. 1--12.

\bibitem{Khojastepour}
M.~A. Khojastepour, K.~Sundaresan, S.~Rangarajan, X.~Zhang, and S.~Barghi,
  ``The case for antenna cancellation for scalable full-duplex wireless
  communications,'' in \emph{Proceedings of the 10th ACM Workshop on Hot Topics
  in Networks}, 2011, pp. 17:1--17:6.

\bibitem{Jain11}
M.~Jain, J.~I. Choi, T.~Kim, D.~Bharadia, S.~Seth, K.~Srinivasan, P.~Levis,
  S.~Katti, and P.~Sinha, ``Practical, real-time, full duplex wireless,'' in
  \emph{17th Annual International Conference on Mobile Computing and
  Networking}, Sept. 2011, pp. 301--312.

\bibitem{Sabharwal_Survey}
A.~Sabharwal, P.~Schniter, D.~Guo, D.~Bliss, S.~Rangarajan, and R.~Wichman,
  ``In-band full-duplex wireless: Challenges and opportunities,'' \emph{IEEE
  Journal on Selected Areas in Communications}, vol.~32, no.~9, pp. 1637--1652,
  Sept 2014.

\bibitem{Bharadia}
D.~Bharadia, E.~McMilin, and S.~Katti, ``Full duplex radios,'' \emph{SIGCOMM
  Comput. Commun. Rev.}, vol.~43, no.~4, pp. 375--386, Aug. 2013.

\bibitem{Day12}
B.~Day, A.~Margetts, D.~Bliss, and P.~Schniter, ``Full-duplex {MIMO} relaying:
  Achievable rates under limited dynamic range,'' \emph{IEEE Journal on
  Selected Areas in Communications}, vol.~30, no.~8, pp. 1541 --1553, Sept.
  2012.

\bibitem{Riihonen11}
T.~Riihonen, S.~Werner, and R.~Wichman, ``Hybrid full-duplex/half-duplex
  relaying with transmit power adaptation,'' \emph{IEEE Transactions on
  Wireless Communications}, vol.~10, no.~9, pp. 3074--3085, Sept 2011.

\bibitem{Ng}
D.~Ng, E.~Lo, and R.~Schober, ``Dynamic resource allocation in {MIMO-OFDMA}
  systems with full-duplex and hybrid relaying,'' \emph{IEEE Transactions on
  Communications}, vol.~60, no.~5, pp. 1291--1304, May 2012.

\bibitem{SImodel}
M.~Duarte, C.~Dick, and A.~Sabharwal, ``Experiment-driven characterization of
  full-duplex wireless systems,'' \emph{IEEE Transactions on Wireless
  Communications}, vol.~11, no.~12, pp. 4296--4307, December 2012.

\bibitem{Rodriguez2}
L.~Jimenez~Rodriguez, N.~Tran, and T.~Le-Ngoc, ``Optimal power allocation and
  capacity of full-duplex af relaying under residual self-interference,''
  \emph{IEEE Wireless Communications Letters}, vol.~3, no.~2, pp. 233--236,
  April 2014.

\bibitem{Sanaz}
S.~Barghi, A.~Khojastepour, K.~Sundaresan, and S.~Rangarajan, ``Characterizing
  the throughput gain of single cell mimo wireless systems with full duplex
  radios,'' in \emph{10th International Symposium on Modeling and Optimization
  in Mobile, Ad Hoc and Wireless Networks}, May 2012, pp. 68--74.

\bibitem{Aggarwal12}
V.~Aggarwal, M.~Duarte, A.~Sabharwal, and N.~K. Shankaranarayanan, ``Full- or
  half-duplex? {A} capacity analysis with bounded radio resources,'' in
  \emph{IEEE Information Theory Workshop}, Sep 2012.

\bibitem{Tse}
D.~Tse and P.~Viswanath, \emph{Fundamentals of Wireless Communication}.\hskip
  1em plus 0.5em minus 0.4em\relax New York, NY, USA: Cambridge University
  Press, 2005.

\bibitem{Ju11}
H.~Ju, X.~Shang, H.~Poor, and D.~Hong, ``Bi-directional use of spatial
  resources and effects of spatial correlation,'' \emph{IEEE Transactions on
  Wireless Communications}, vol.~10, no.~10, pp. 3368--3379, October 2011.

\bibitem{estimation}
D.~Kim, H.~Ju, S.~Park, and D.~Hong, ``Effects of channel estimation error on
  full-duplex two-way networks,'' \emph{IEEE Transactions on Vehicular
  Technology}, vol.~62, no.~9, pp. 4666--4672, Nov 2013.

\bibitem{Arifin}
A.~Arifin and T.~Ohtsuki, ``Outage probability analysis in bidirectional
  full-duplex siso system with self-interference,'' in \emph{Asia-Pacific
  Conference on Communications}, Oct 2014, pp. 6--8.

\bibitem{Comtel}
K.~Akcapinar and O.~Gurbuz, ``Full-duplex bidirectional communication under
  self-interference,'' in \emph{13th International Conference on
  Telecommunications}, July 2015, pp. 1--7.

\bibitem{FDMIMO}
D.~Bharadia and S.~Katti, ``Full duplex mimo radios,'' in \emph{Proceedings of
  the 11th USENIX Conference on Networked Systems Design and Implementation},
  2014, pp. 359--372.

\bibitem{Pashazadeh}
M.~Pashazadeh and F.~Tabataba, ``Performance analysis of one-way relay networks
  with channel estimation errors and loop-back interference,'' in \emph{23rd
  Iranian Conference on Electrical Engineering}, May 2015, pp. 432--437.

\bibitem{goyal_improving}
S.~Goyal, P.~Liu, S.~Panwar, R.~Difazio, R.~Yang, J.~Li, and E.~Bala,
  ``Improving small cell capacity with common-carrier full duplex radios,'' in
  \emph{IEEE International Conference on Communications}, June 2014, pp.
  4987--4993.

\bibitem{Alves}
H.~Alves, D.~Benevides~da Costa, R.~Demo~Souza, and M.~Latva-aho, ``On the
  performance of two-way half-duplex and one-way full-duplex relaying,'' in
  \emph{IEEE 14th Workshop on Signal Processing Advances in Wireless
  Communications}, June 2013, pp. 56--60.

\bibitem{FDR_Survey}
G.~Liu, F.~Yu, H.~Ji, V.~Leung, and X.~Li, ``In-band full-duplex relaying: A
  survey, research issues and challenges,'' \emph{IEEE Communications Surveys
  \& Tutorials}, vol.~17, no.~2, pp. 500--524, Secondquarter 2015.

\bibitem{Goldsmith}
A.~Goldsmith, \emph{Wireless Communications}.\hskip 1em plus 0.5em minus
  0.4em\relax New York, NY, USA: Cambridge University Press, 2005.

\bibitem{Gaussian_Noise1}
S.~Diggavi and T.~Cover, ``The worst additive noise under a covariance
  constraint,'' \emph{IEEE Transactions on Information Theory}, vol.~47, no.~7,
  pp. 3072--3081, Nov 2001.

\bibitem{Gaussian_Noise2}
I.~Shomorony and A.~Avestimehr, ``Worst-case additive noise in wireless
  networks,'' \emph{IEEE Transactions on Information Theory}, vol.~59, no.~6,
  pp. 3833--3847, June 2013.

\bibitem{shannon1961}
C.~E. Shannon, ``Two-way communication channels,'' in \emph{Proceedings of the
  Fourth Berkeley Symposium on Mathematical Statistics and Probability, Volume
  1: Contributions to the Theory of Statistics}, 1961, pp. 611--644.

\bibitem{DaF}
C.~Esli and A.~Wittneben, ``One- and two-way decode-and-forward relaying for
  wireless multiuser{ MIMO} networks,'' in \emph{IEEE Global Telecommunications
  Conference}, Nov 2008, pp. 1--6.

\bibitem{4418498}
T.~Oechtering, C.~Schnurr, I.~Bjelakovic, and H.~Boche, ``Broadcast capacity
  region of two-phase bidirectional relaying,'' \emph{IEEE Transactions on
  Information Theory}, vol.~54, no.~1, pp. 454--458, 2008.

\bibitem{katti_embracing}
S.~Katti, S.~Gollakota, and D.~Katabi, ``Embracing wireless interference:
  Analog network coding,'' \emph{SIGCOMM Comput. Commun. Rev.}, vol.~37, no.~4,
  pp. 397--408, Aug. 2007.

\bibitem{Zhang06}
S.~Zhang, S.~C. Liew, and P.~P. Lam, ``Hot topic: Physical-layer network
  coding,'' in \emph{Proceedings of the 12th Annual International Conference on
  Mobile Computing and Networking}, 2006, pp. 358--365.

\bibitem{3models}
J.~I. Choi, S.~Hong, M.~Jain, S.~Katti, P.~Levis, and J.~Mehlman, ``Beyond full
  duplex wireless,'' in \emph{Forty Sixth Asilomar Conference on Signals,
  Systems and Computers}, Nov 2012, pp. 40--44.

\bibitem{softnull}
E.~Everett, C.~Shepard, L.~Zhong, and A.~Sabharwal, ``Softnull: Many-antenna
  full-duplex wireless via digital beamforming,'' \emph{IEEE Transactions on
  Wireless Communications}, vol.~15, no.~12, pp. 8077--8092, Dec 2016.

\bibitem{Shojaeifard}
A.~Shojaeifard, K.~Wong, M.~D. Renzo, G.~Zheng, K.~A. Hamdi, and J.~Tang,
  ``Self-interference in full-duplex multi-user {MIMO} channels,'' \emph{CoRR},
  vol. abs/1701.00277, 2017.

\bibitem{Jafar}
S.~A. Jafar and M.~J. Fakhereddin, ``Degrees of freedom for the {MIMO}
  interference channel,'' \emph{IEEE Transactions on Information Theory},
  vol.~53, no.~7, pp. 2637--2642, July 2007.

\bibitem{knopp}
R.~Knopp, ``Two-way wireless communication via a relay station,'' in
  \emph{GDR-ISIS meeting}, Mar. 2007.

\end{thebibliography}

\end{document}